\documentclass[aps,prc,twocolumn,showpacs,nofootinbib]{revtex4-1}
\usepackage{graphicx}   %       for graphics
\graphicspath{{/home/tom/Physik/FOPI/figures/}}
\usepackage{latexsym}   %       for special symbols
\usepackage{enumerate}
\usepackage{rotating,booktabs,multirow}
\begin{document}
%============================================================================
\title{$\Delta$ resonances in Ca+Ca, Ni+Ni and Au+Au reactions from 1 AGeV to 2 AGeV: Consistency between yields, mass shifts and decoupling temperatures}
\author{Tom Reichert$^{1,4}$,
Paula~Hillmann$^{1,2,3,4}$,
Marcus~Bleicher$^{1,2,3,4}$}

\affiliation{$^1$ Institut f\"ur Theoretische Physik, Goethe Universit\"at Frankfurt, Max-von-Laue-Strasse 1, D-60438 Frankfurt am Main, Germany}
\affiliation{$^2$ GSI Helmholtzzentrum f\"ur Schwerionenforschung GmbH, Planckstr. 1, 64291 Darmstadt , Germany}
\affiliation{$^3$ John von Neumann-Institut f\"ur Computing, Forschungzentrum J\"ulich,
52425 J\"ulich, Germany}
\affiliation{$^4$ Helmholtz Research Academy Hesse for FAIR, Campus Frankfurt, Max-von-Laue-Str. 12, 60438 Frankfurt,  Germany}

\begin{abstract}
	The Ultra-relativistic Quantum Molecular Dynamics (UrQMD) transport approach is used to calculate $\Delta$(1232) yields in Ca+Ca, Ni+Ni and Au+Au collisions between 1~AGeV and 2~AGeV. We compare and validate two different methods to extract the yields of $\Delta$(1232) resonances in such low energy nuclear collisions: Firstly, the $\pi^-$ spectra at low $p_\mathrm{T}$ are used to infer the $\Delta$(1232) yield in A+A collisions, a method employed by the GSI/FOPI experiment. Secondly, we employ the invariant mass method used by the HADES collaboration, which has recently reported data in the $\Delta^{++}\rightarrow\pi^++p$ channel. We show that both methods are compatible with each other. Then we use the $\Delta/nucleon$ ratio to extract the kinetic decoupling temperatures of the $\Delta$(1232) resonances. We find that the extracted temperatures are consistent with the predicted mass shift of the $\Delta$ resonance and the freeze-out parameters estimated from complementary studies (blast wave fits, coarse graining).
\end{abstract}

\maketitle
\section{Introduction}
Heavy ion collisions, carried out in today's largest particle accelerators, provide excellent opportunities to study nuclear and sub-nuclear matter at extreme conditions. With increasing energy, the possibility to produce novel and exotic states of matter becomes accessible. In the GSI/NICA/FAIR energy regime covering the collision energy range from 1-20~AGeV the exploration of highest densities is in the focus of the experimental programs, opening the route to explore the type of matter present in the interior of neutron stars. At higher energies, nearly net-baryon free matter at highest temperatures is studied. Conclusions about the properties of subatomic matter are drawn from particle distributions, e.g. by PHENIX \cite{Adler:2003cb}, STAR \cite{Agakishiev:2011ar}, ALICE \cite{Abelev:2013vea} and CMS \cite{Chatrchyan:2012qb}. Due to the explosive nature of heavy ion reactions, the time scales of the reactions do not allow for a direct observation of the reaction zone, but one needs to infer the properties of the created QCD-matter at the different stages of the reaction indirectly, e.g. via flow measurements or penetrating probes. Nevertheless it remains difficult and often ambiguous to pin down specific values for the parameters (temperature, density, expansion velocity, transport properties) of the emission source. In this paper we want to answer one of these questions, namely the value of the decoupling temperature, with the help of the Delta resonance.

Historically the $\Delta$(1232) has long been of major interest as its discovery lead to the concept of color charge to maintain the Pauli principle in the $\Delta^{++}$ and the $\Delta^{-}$ states. For our current exploration, the $\Delta$(1232) is a prime candidate because due to its relatively low mass it is the most abundantly created baryonic resonance. Apart from leptonic decays with very small branching ratios, the $\Delta$(1232) decays always into a $\pi$ and a $N$ and is therefore rather easy to produce and to measure. As was shown in \cite{Reichert:2019lny,Reichert:2019zab} its spectral function is linked to the temperature of the system at the decoupling surface. Interestingly, a complementary method based on the chemical yields of short-lived resonances to extract the decoupling properties was developed shortly after this finding \cite{Motornenko:2019jha} and will also be used here to analyze and compare the model results. 

The experimental measurement of a decaying $\Delta$(1232) resonance can be accomplished in several ways. Today, the most common way is to detect it in the invariant mass distribution of charged $\pi+N$ pairs. Measurements using this method have recently been reported by the HADES collaboration at GSI \cite{Kornakov:talk2019}. Nowadays, the 2-particle (or even n-particle) invariant mass reconstruction is the method of choice, if sufficient statistics and detector resolution permit its use. Sometimes, however, also simpler methods (using only 1-particle information) can be employed, e.g. high $p_{\rm T}$ single electrons can be seen as proxies for $D$-mesons \cite{Gossiaux:2008jv}. In the case of the $\Delta$(1232) a simplified experimental analysis can also be performed for nucleus-nucleus reactions at low energies \cite{Larionov:1999iw}. Here, one uses the tight correlation between the pion yield and the $\Delta$(1232) yield: The Delta, having a lifetime of $\sim$~1~fm and being the most dominant source of pions at low energies allows to detect $\Delta$ resonances via the $\pi$ spectra. The observed transverse momentum distribution of pions has two contributions, direct pions and and additional pions emerging from decaying $\Delta$(1232). These pions originating from Delta decays populate low transverse momenta and dominate the spectrum below $p_T\approx 400$~MeV. Thus, low $p_\mathrm{T}$ enhancement, together with theoretical calculations, can be used to investigate the $\Delta$(1232) resonance without invariant mass reconstruction. Pioneering results employing this method have been published for Si+Au reactions at 14.6~AGeV at the AGS \cite{Brown:1991en}, by the E814 collaboration in Si+Al, Si+Al and Si+Pb reactions at 14.6~AGeV at the AGS \cite{Barrette:1992kh}, and by the FOPI collaboration in Ni+Ni collisions between 1~AGeV and 2~AGeV at GSI \cite{Hong:1997ka}. The obtained $\Delta/$nucleon ratios \cite{Hong:1997ka} can then be used to estimate the freeze-out temperature at the decoupling hyper-surface of the $\Delta$ using the hadrochemical equilibrium model \cite{BraunMunzinger:1994xr}. 

In this paper we will be a) providing a consistency check between different methods to estimate the abundance of $\Delta$ resonances to connect the FOPI data \cite{Hong:1997ka} to the recently measured HADES data on Delta resonances \cite{Kornakov:talk2019,Adamczewski-Musch:2020vrg} and b) extracting the decoupling temperature of the $\Delta$ from a chemical analysis of the $\Delta/nucleon$ yield ratio to compare to kinetic freeze-out studies and apparent mass shifts. To this aim, we will employ the UrQMD model (v3.4) \cite{Bass:1998ca,Bleicher:1999xi} which has been used to explore a wide range of reactions in this energy regime, e.g. flow \cite{Hillmann:2018nmd,Hillmann:2019cfr,Hillmann:2019wlt}, strangeness \cite{Steinheimer:2015sha}, clusters \cite{Sombun:2018yqh}, and di-leptons \cite{Endres:2015fna}. The different techniques to extract $\Delta$ yields discussed above are then applied to Ca+Ca (Ar+KCl) collisions at 1.76~AGeV and Au+Au collisions at 1.23~AGeV to predict the relative $\Delta$(1232) abundance at the HADES energies. For alternative studies we refer to \cite{Hartnack:1997ez,Ko:1999qh,Larionov:1999iw,Larionov:2001va}

\section{The UrQMD model}
We employ the Ultra relativistic Quantum Molecular Dynamics (UrQMD) transport model in its most recent version (v3.4) \cite{Bass:1998ca,Bleicher:1999xi} in cascade mode. It is based on the covariant propagation of hadrons and their interactions by elastic and/or inelastic collisions. Relevant cross sections are taken, if available, from experimental data or derived from effective models. UrQMD includes mesonic and baryonic resonances up to masses of 2~GeV and has a longstanding history to reproduce hadron yields and spectra and forecast new phenomena. In case of the GSI energy regime investigated in the present study, UrQMD has already been proven to describe various observables. For recent results, we refer the reader to explorations of Au+Au collisions at 1.23~AGeV \cite{Steinheimer:2015sha,Endres:2015fna,Hillmann:2018nmd,Hillmann:2019wlt,Hillmann:2019cfr}, and Ca+Ca collisions at 1.76~AGeV \cite{Steinheimer:2016vzu} or in Ar+KCl collisions at 1.76~AGeV \cite{Steinheimer:2015sha,Endres:2015fna} which are described very well. 

A major advantage of a transport simulation is the opportunity to analyze the full time evolution of each event. This means, in contrast to the two analysis methods available to experimentalists, we can track down each decaying $\Delta$(1232) resonance and follow its individual daughter particles until their next interaction. This allows to define microscopically, if a given Delta might be observable or not. E.g., if both daughters re-scatter only elastically (probably many times) and do not re-scatter inelastically then the decaying $\Delta$ is considered as reconstructable. This method has already proven to be reliable and is described in detail in \cite{Bleicher:2002dm,Reichert:2019lny,Reichert:2019zab}.  

\section{Particle spectra}
To benchmark the model and the reconstruction methods we calculate Ca+Ca, Ni+Ni , and Au+Au reactions with the UrQMD model. The uppermost 4\% of the total cross section are selected via the geometrical interpretation of the cross section using a sharp sphere approximation. This translates to impact parameters of $b_{\rm max}=1.6$~fm (Ca+Ca), $b_{\rm max}=1.9$~fm (Ni+Ni) and $b_{\rm max}=2.8$~fm (Au+Au) respectively. 

\begin{figure} [t!hb]
	\includegraphics[width=\columnwidth]{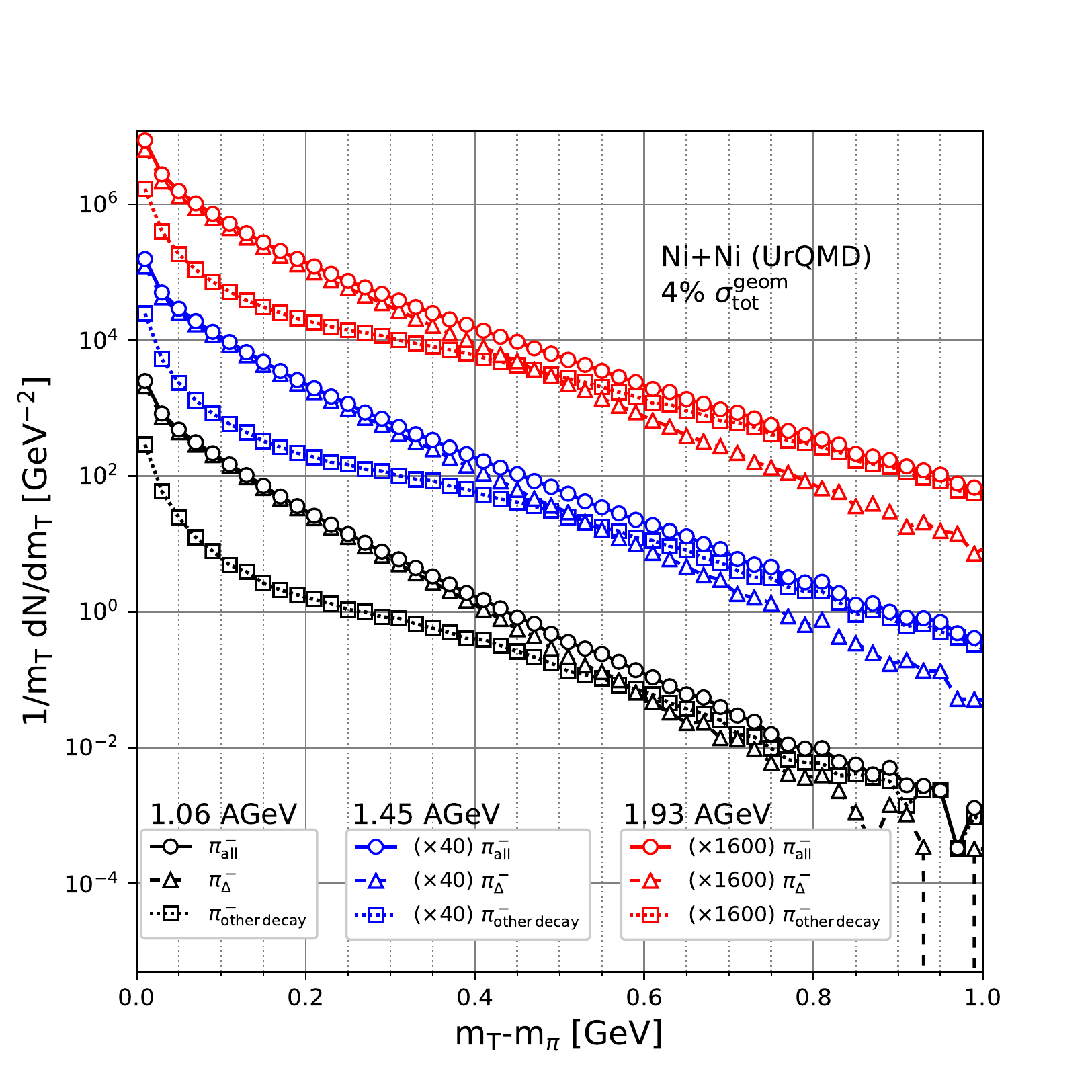}
	\caption{[Color online] Transverse mass distributions of all $\pi^-$ (circles), of $\pi^-$ from decaying $\Delta$s (triangles) and of $\pi^-$ from other decaying resonances (squares) in central Ni+Ni collisions at 1.06~AGeV (black), 1.45~AGeV (blue) and 1.93~AGeV (red) from UrQMD.}
	\label{dndmt}
\end{figure}

\begin{figure} [t!hb]
	\includegraphics[width=\columnwidth]{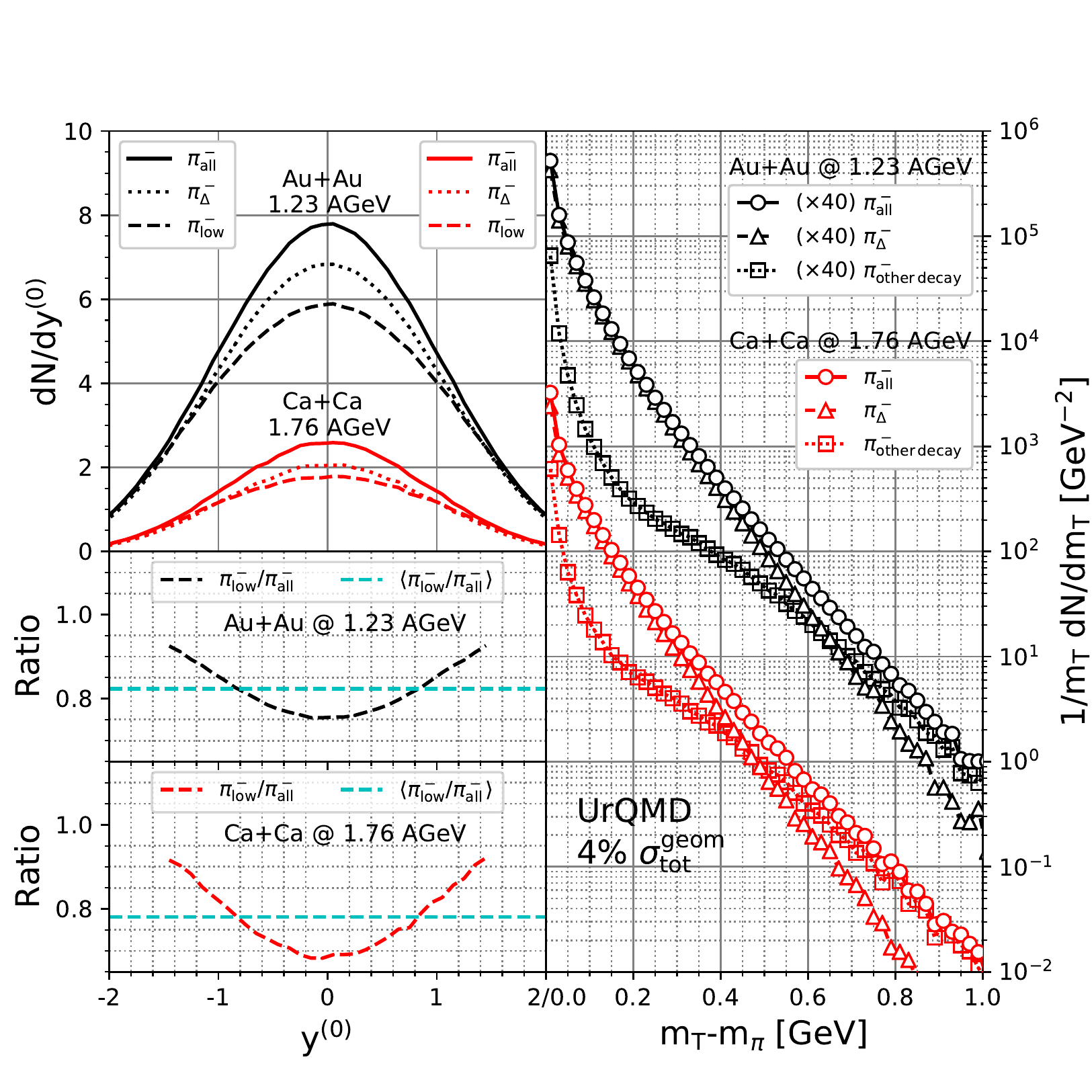}
	\caption{[Color online] Rapidity distributions (upper left panel) of $\pi^-$ at central Ca+Ca collisions at 1.76~AGeV (red) and central Au+Au collisions at 1.23~AGeV (black) as well as their ratios (lower left panels). The transverse mass distributions (right panel) are shown for all pions (circles), pions from $\Delta$ decays (triangles) and pions from other resonance decays (squares) from the Ca+Ca system (red) and from the Au+Au system (black).}
	\label{dndy2}
\end{figure}

Let us start with the transverse mass spectra in Fig. \ref{dndmt} for the Ni+Ni systems at three different beam energies (1.06 AGeV, 1.45 AGeV, 1.93 AGeV). (The transverse mass spectra for Ca+Ca and Au+Au are shown in the right panel of Fig. \ref{dndy2} and will be discussed in more detail later.) For each energy, we split the total $\pi^-$ yield (circles) into a contribution of $\pi^-$ originating directly from $\Delta$ decays (triangles) and $\pi^-$ originating from the decay of other resonances (squares). The influence of the $\Delta$ decay is clearly visible  for all three shown energies and dominates the pion yield up to transverse masses of $\approx0.4$~GeV. Assuming $\Delta$ resonances with nominal mass and at rest, it is clear that the transverse mass of an emitted pion in the local rest frame of the Delta amounts to a maximal value of $\approx$~0.3~GeV matching the observed upper $p_\mathrm{T}$ limit in the Lorentz-boosted frame. Thus the $\Delta$ is the major contributor to the pion spectrum at low $p_\mathrm{T}$. Pions at higher $p_\mathrm{T}$ are created via  the decay of mesonic resonances or heavier baryonic resonances and populate the (strongly suppressed) high $p_{\rm T}$ tail of the distribution. 

\begin{figure} [t!hb]
	\includegraphics[width=\columnwidth]{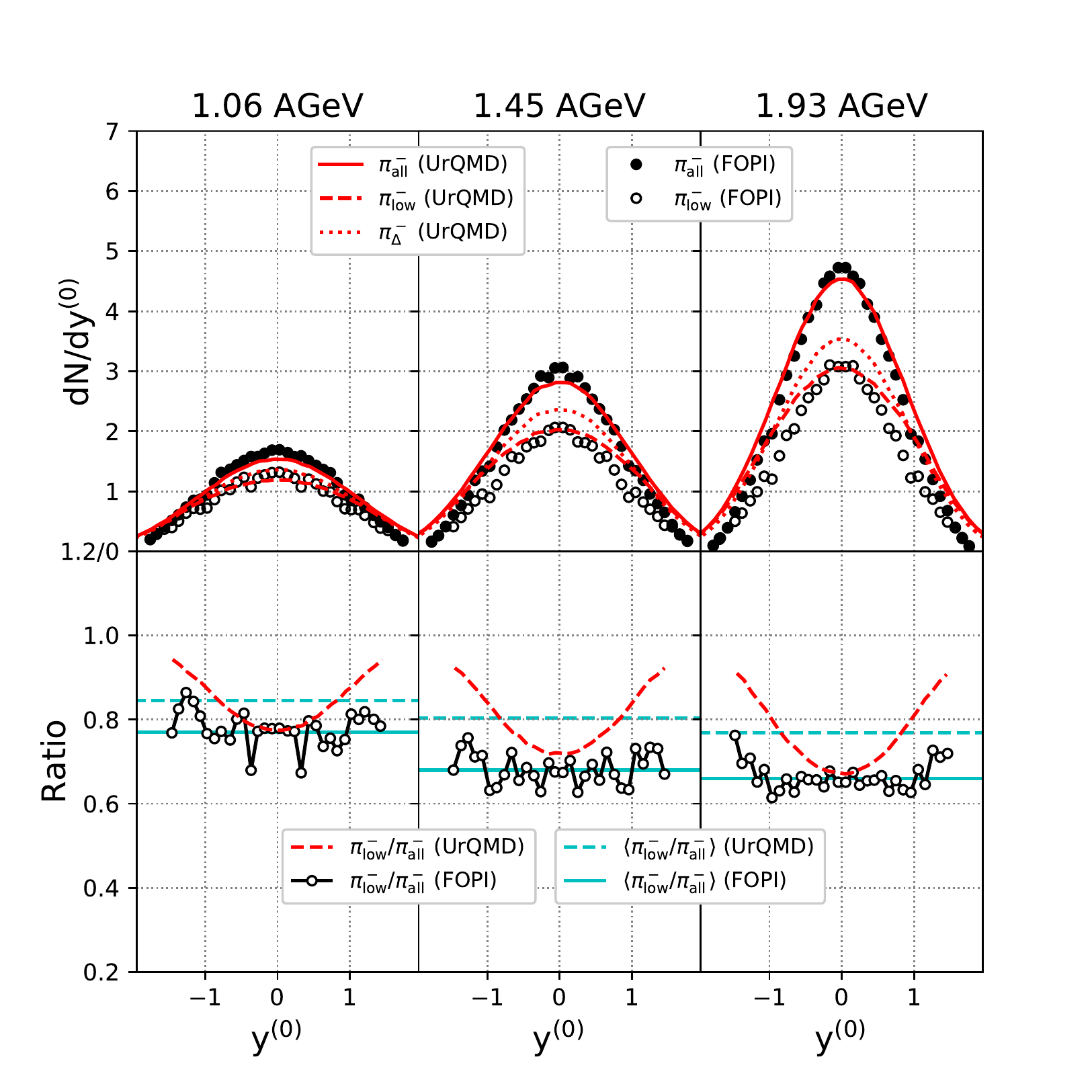}
	\caption{[Color online] Rapidity distributions (upper panel) of all $\pi^-$ (red solid lines), of $\pi^-$ at low $p_{\rm T}$ (red dashed lines) and of $\pi^-$ from decaying $\Delta$s (red dotted lines) at central Ni+Ni collisions at 1.06~AGeV (left), 1.45~AGeV (middle) and 1.93~AGeV (right) and the ratio of low $p_{\rm T}$ pions to all $\pi^-$ (lower panel) from UrQMD. Experimental data points for all $\pi^-$ (full black circles) and for $\pi^-$ at low $p_{\rm T}$ (empty black circles) are taken from \cite{Hong:1997ka}.}
	\label{dndy}
\end{figure}

After setting the stage, we can now investigate the rapidity distribution of pions in Ni+Ni collisions at the same three energies as above as shown in Fig. \ref{dndy} (upper panel). The beam energy increases from left to right from 1.06~AGeV (left panel), over 1.45~AGeV (middle panel) to 1.93~AGeV (right panel). Note that the  distributions are scaled to normalized rapidity $y^{(0)}=y/y_{\mathrm{c.m.s}}-1$ to take care of the different collision energies. For each energy, we show the rapidity densities of all $\pi^-$ (red solid line),  $\pi^-$ with $p_\mathrm{T}\leq0.3$~GeV (dubbed $\pi_{\rm low}$, red dashed line) and of $\pi^-$ originating from $\Delta$(1232) decays (dubbed $\pi_{\Delta}$, red dotted line). The calculations are compared to the FOPI data taken from \cite{Hong:1997ka} showing all $\pi^-$ (full black circles) and low $p_\mathrm{T}$ $\pi^-$ (empty black circles).  

As speculated above, the model calculation also shows clearly that the low $p_\mathrm{T}$ pions provide an excellent approximation (within 10\%) for the pions coming from the decay of a $\Delta$(1232) as seen from the comparison of the dotted and dashed lines. 

The lower panel of Fig. \ref{dndy} shows the ratio of the low $p_\mathrm{T}$ $\pi^-$ to all $\pi^-$ (UrQMD: red dashed line, FOPI: black line with circles) as a function of normalized rapidity. Generally this ratio shows a mild rapidity dependence, both in the simulation and the data. Generally, the relative abundance of low $p_{\rm T}$ pions decreases with increasing energy due to additional production channels and radial flow. To simplify the extraction of the Delta yield (and to follow the FOPI analysis), we approximate the rapidity dependent ratio, by its average $\langle\pi^-_{\mathrm{low}}/\pi^-_\mathrm{all}\rangle$ (cyan dashed line) calculated for $|y^{(0)}|\le1.5$ as done in \cite{Hong:1997ka} which is a good approximation in the 10\% level.

Let us now use this method to analyze the recent HADES data with respect to the Delta yield obtained in different systems and at different energies using the same method. The right part of Fig. \ref{dndy2} analyzes the transverse mass distributions of pions for the Ca+Ca system at 1.76~AGeV (red) and for the Au+Au system at 1.23~AGeV (black) as obtained from the UrQMD simulation. Again, we split the total $\pi^-$ yield (circles) into a contribution of $\pi^-$ originating from $\Delta$ decays (triangles) and $\pi^-$ originating from the decay of other resonances (squares). The contribution of the $\Delta$ decay is clearly visible for both systems and dominates the pion yield up to transverse masses of $\approx0.4$~GeV. In the upper left panel of Fig. \ref{dndy2} we show the rapidity distributions of pions for the same systems.  The $p_{\rm T}$-integrated $\pi^-$ distributions are shown as solid lines, the low transverse momentum $\pi^-$, i.e. $p_\mathrm{T}\leq0.3$~GeV are depicted by dashed lines and the $\pi^-$ originating from $\Delta$(1232) decays are indicated by dotted lines. As before, we observe that the low $p_{\rm T}$ pions provide a good proxy  for the pions stemming from the decay of Delta resonances (within 20\% for the Au+Au case). The ratio of the low $p_\mathrm{T}$ $\pi^-$ to all $\pi^-$ as a function of rapidity is shown in the lower left panels of Fig. \ref{dndy2}, see legend. As discussed above, we observe again that the importance of low transverse momenta decreases with increasing energy. 

\section{Comparison of resonance abundances}
We can now estimate the Delta yields and the fraction of Delta resonances from all participating baryons. We do this in two different ways: Firstly, we follow the FOPI approach to estimate the $\Delta$(1232) abundance. i.e., we multiply the number of pions with the average $\langle\pi^-_\mathrm{low}$/$\pi^-_\mathrm{all}\rangle$ ratio at freeze-out and employ a scaling factor, taken from the isobar model \cite{Stock:1985xe} to account for the isospin asymmetry, as depicted in Eq. \ref{isobar}: %
\begin{figure} [thb]
	\includegraphics[width=\columnwidth]{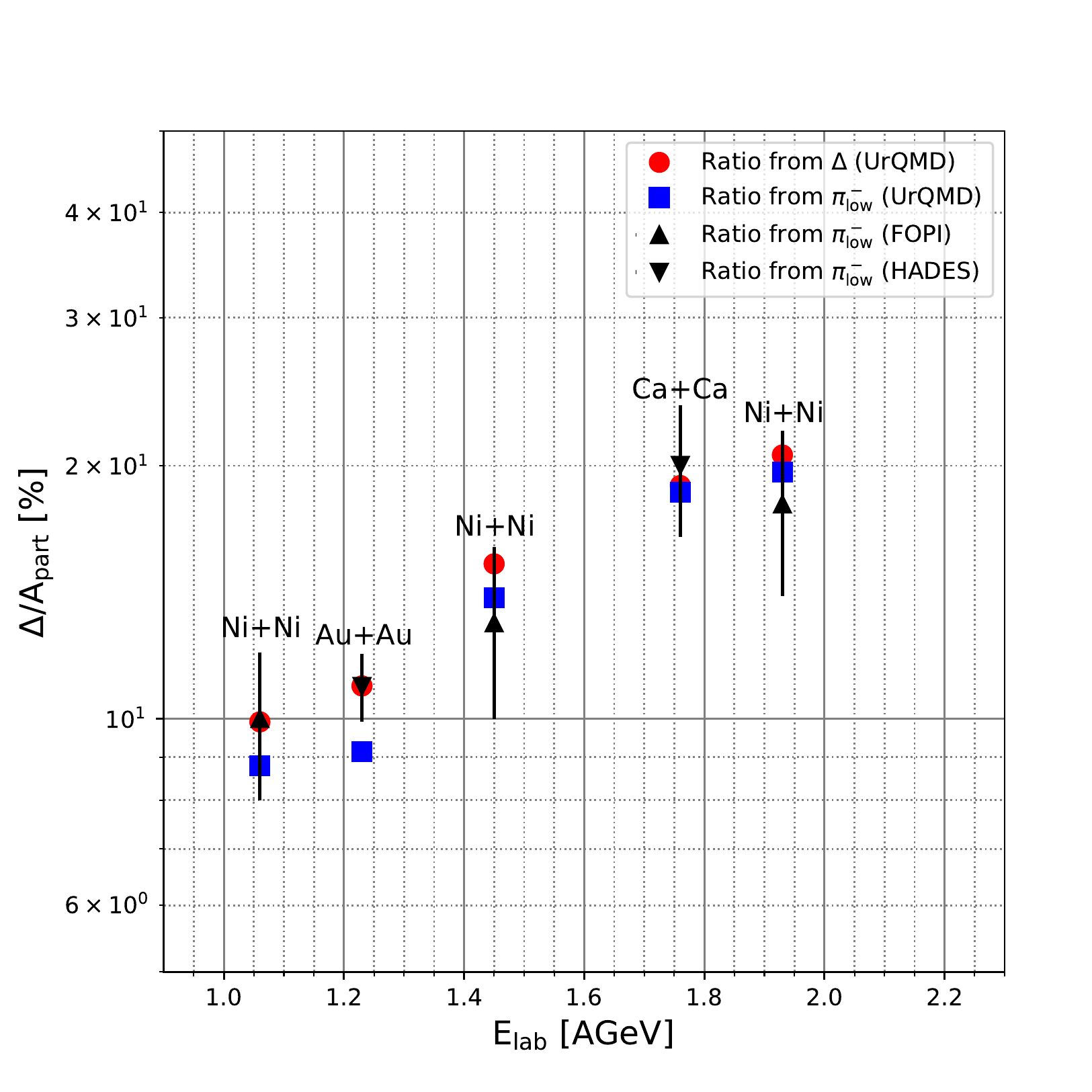}
	\caption{[Color online] Relative amount of Delta resonances as a function of E$_\mathrm{lab}$. The red circles show the UrQMD results of the microscopic calculation of the Delta yield, while the blue squares show the Delta yield extracted via the low $p_{\rm T}$ pion yield. The experimental data points of FOPI are shown as black triangles \cite{Hong:1997ka}, the HADES result for Au+Au and Ca+Ca reactions is shown as inverted triangle \cite{Adamczewski-Musch:2020vrg,Tlusty:2009dk}.}
	\label{ratio_elab}
\end{figure}
\begin{equation}\label{isobar}
	n(\Delta)\approx n(\pi^-)\times f_\mathrm{isobar}\times\left\langle\frac{\pi^-_\mathrm{low}}{\pi^-_\mathrm{all}}\right\rangle.
\end{equation}
The $f_\mathrm{isobar}$ factor has the numerical values 2.84 ($^{58}$Ni+$^{58}$Ni), 2.24 ($^{197}$Au+$^{197}$Au) and 3.0 ($^{40}$Ca+$^{40}$Ca). The obtained abundance of Delta resonances serves as input to calculate the ratio of excited nucleons to nucleons as shown by Eq. \ref{ratio}:
\begin{equation}\label{ratio}
	R=\frac{n(\Delta)}{A_{\rm participant}},
\end{equation}
where $A_{\rm participant}$ includes all ground state nucleons and all decay nucleons from resonances, i.e. $A_{\rm participant}\approx n(\Delta)+n(nucleon)$ at the Delta freeze-out surface.

We are now in the position to compare the different methods to extract Delta yields for different systems and collision energies.
Fig. \ref{ratio_elab} shows the fraction of $\Delta$(1232) resonances to all participating nucleons in dependence of the beam energy. Two model calculations are compared to the data by the FOPI collaboration: The red circles show the results using the microscopically reconstructed true Delta resonances from UrQMD, the blue squares show the UrQMD results using only low $p_{\rm T}$ pions. We clearly observe that both approaches yield similar results. In addition, one observes that both approaches are fully compatible with the experimental estimates provided by the FOPI collaboration (black triangles with error bars) \cite{Hong:1997ka}. It is especially interesting to note that also the new HADES data\footnote[1]{We have used the new HADES data on pion production in Au+Au reactions at 1.23 AGeV \cite{Adamczewski-Musch:2020vrg} to extract the Delta resonance yield using the same method as used by the FOPI collaboration. I.e. we use eq. \ref{isobar} with $n(\pi^-)=17.1\pm 0.9$, $\left\langle\pi^-_\mathrm{low}/\pi^-_\mathrm{all}\right\rangle_{y_\mathrm{cm}}=0.86$. The number of participants for central collisions was given as $303\pm 12$.}%
 (Au+Au \cite{Adamczewski-Musch:2020vrg}, Ca+Ca \cite{Tlusty:2009dk} shown by an inverted triangle with error bars) is consistent with the energy dependence systematics spanned by the FOPI data. After consistency between the methods and the experiments has been established, we can now interpret the results in a straightforward fashion: The fraction of resonances is clearly energy (i.e. temperature) dependent and increases from 10\% to 20\% when going from 1 AGeV to 2 AGeV beam energy. It is also clear that the resonance fraction is not (or only very weakly) system size dependent, which is compatible with estimates from thermal models that do not see a sizable centrality dependence of freeze-out parameters.

\section{Temperature extraction}
Finally, we can use the obtained $\Delta$/($\Delta$+nucleons) ratios to calculate the temperature of the decoupling hyper-surface of the observable $\Delta$ resonances. To this aim we use the thermal model in Boltzmann approximation:
\begin{equation}
n_i= (2s_i+1)(2l_i+1)V\frac{T^3e^{\mu/T}}{2\pi^2}\frac{m_i^2}{T^2}K_2\left(\frac{m_i}{T}\right).
\end{equation}
Here, $s_i$ is the spin, $l_i$ the iso-spin of the hadron, $m_i$ is its mass, while $T$ is the temperature, $V$ the volume and $\mu$ is the chemical potential, $K_2$ denotes the Bessel function. Note that $n_{\rm nucleon}$ includes the decay contributions from the Deltas. Table \ref{Temp} shows the estimated temperatures calculated from the Delta yields with the two different methods. The simulation results are compared to the FOPI results \cite{Hong:1997ka}. The 1$\mathrm{st}$ column shows the energy and the system while the estimated temperatures from the corresponding methods are shown from column 2 to column 5. The 2$\mathrm{nd}$ column shows the temperature results obtained by using the microscopically reconstructed $\Delta$(1232) yield from UrQMD, the 3$\mathrm{rd}$ column shows the temperature calculated with the low $p_{\rm T}$ method from UrQMD and both are compared to the FOPI results \cite{Hong:1997ka} using the low $p_{\rm T}$ pion method in the 4$\mathrm{th}$ column and using a radial flow analysis in column 5. 
\begin{table} [thb]
	\begin{tabular}{l|c|c|c|c}
		E$_\mathrm{lab}$ [AGeV]
		&\multicolumn{4}{c}{T [MeV]} \\ \hline \hline
		(System)&$T_\Delta^\mathrm{UrQMD}$ & $T_{\pi_\mathrm{low}}^\mathrm{UrQMD}$ & $T_\mathrm{\Delta}^\mathrm{FOPI}$& $T_\mathrm{Flow}^\mathrm{FOPI}$ \\
		\hline
		1.06 (Ni+Ni) & 74 & 72 & 75$\pm$5 & 79$\pm$10 \\
		1.23 (Au+Au) & 76 & 72 & -- & --\\
		1.45 (Ni+Ni) & 85 & 82 & 80$\pm$7 & 84$\pm$10\\
		1.76 (Ca+Ca) & 92 & 91 & -- & --\\
		1.93 (Ni+Ni) & 95 & 93 & 89$\pm$9 & 92$\pm$12\\
	\end{tabular}
	\caption{Temperatures obtained from the $\Delta$/($A_{\rm part}$) ratio of UrQMD using the thermal model. The FOPI results are taken from \cite{Hong:1997ka}. }
	\label{Temp}      
\end{table}

As can be seen from Tab. \ref{Temp}, the estimated temperatures of the different methods used are in good agreement with each other and the measurements by the FOPI collaboration. With increasing collision energy the temperature rises from $\approx$~73~MeV to $\approx$~94~MeV which is in line with \cite{Rode:2018hlj}. The temperature extracted at 1.23~AGeV in the Au+Au system re-confirms the value obtained by the analysis of the predicted and measured mass shift of the $\Delta$(1232) which suggested a freeze-out temperature of 81~MeV and a mass shift of $-47$ MeV \cite{Reichert:2019lny,Reichert:2019zab}. Based on the extracted temperatures, we predict kinetic mass shifts of $-56$~MeV (Ni+Ni at 1.06~AGeV), $-55$~MeV (Au+Au at 1.23~AGeV), $-49$~MeV (Ni+Ni at 1.45~AGeV), $-45$~MeV (Ca+Ca at 1.76~AGeV) and $-43$~MeV (Ni+Ni at 1.93~AGeV) using the BW$\times$PS formula described in \cite{Reichert:2019lny,Reichert:2019zab} for the different GSI energies. These mass shifts seem to be confirmed by the estimates provided in \cite{Eskef:2001qg}.

\section{Summary}
In this article we have re-analyzed FOPI data tackling the reconstruction of the $\Delta$(1232) yield from the measured pion yield in Ni+Ni collisions between 1~AGeV and 2~AGeV. The method uses the fact that pions at low transverse momenta originate dominantly from the decay of $\Delta$(1232) resonances. Thus, the pion yield can be scaled by the ratio of low $p_\mathrm{T}$ pions to all pions and the isospin asymmetry factor of the system to obtain the $\Delta$(1232) yield. We compared this method to the reconstruction of the detectable $\Delta$(1232)s by following the daughter particles of each decaying Delta resonance in the simulation. We furthermore compared both methods to predict the relative $\Delta$(1232) abundance for the HADES experiment in Ca+Ca and Au+Au collisions. The results are in very good agreement confirming that both methods can be used to obtain Delta yields in the investigated energy regime. Finally we used the thermal model to estimate the decoupling temperature of the $\Delta$(1232)s. The mean values for the kinetic freeze-out temperature increase from 73~MeV to 94~MeV. The extracted temperature value for Au+Au and Ni+Ni re-confirms our previous estimates for the freeze-out temperature of the Delta resonances and supports our previous finding of a kinematic mass shift related to the freeze-out temperature.

\begin{acknowledgements}
This work was supported by Deutscher Akademischer Austauschdienst (DAAD), Helmholtz Forschungsakademie Hessen (HFHF), and in the framework of COST Action CA15213 THOR. Computational resources were provided by the Center for Scientific Computing (CSC) of the Goethe University.
\end{acknowledgements}

%%%%%%%%%%%%%%%%%%%%%%%%%%%%%%%%%%%%%%%%%%%%%%%%%%%%%%%%%%%%%%%%%%%%%%%%%%%%%%%

%%%%%%%%%%%%%%%%%%%%%%%%%%%%%%%%%%%%%%%%%%%%%%%%%%%%%%%%%%%%%%%%%%%%%%%%%%%%%%% 

\begin{thebibliography}{100}
%\cite{Adler:2003cb}
\bibitem{Adler:2003cb} 
S.~S.~Adler {\it et al.} [PHENIX Collaboration],
%``Identified charged particle spectra and yields in Au+Au collisions at S(NN)**1/2 = 200-GeV,''
Phys.\ Rev.\ C {\bf 69}, 034909 (2004)
doi:10.1103/PhysRevC.69.034909
[nucl-ex/0307022].
%%CITATION = doi:10.1103/PhysRevC.69.034909;%%
%881 citations counted in INSPIRE as of 15 Feb 2020	

%\cite{Agakishiev:2011ar}
\bibitem{Agakishiev:2011ar} 
G.~Agakishiev {\it et al.} [STAR Collaboration],
%``Strangeness Enhancement in Cu+Cu and Au+Au Collisions at $\sqrt{s_{NN}} = 200$ GeV,''
Phys.\ Rev.\ Lett.\  {\bf 108}, 072301 (2012)
doi:10.1103/PhysRevLett.108.072301
[arXiv:1107.2955 [nucl-ex]].
%%CITATION = doi:10.1103/PhysRevLett.108.072301;%%
%85 citations counted in INSPIRE as of 15 Feb 2020

%\cite{Abelev:2013vea}
\bibitem{Abelev:2013vea} 
B.~Abelev {\it et al.} [ALICE Collaboration],
%``Centrality dependence of $\pi$, K, p production in Pb-Pb collisions at $\sqrt{s_{NN}}$ = 2.76 TeV,''
Phys.\ Rev.\ C {\bf 88}, 044910 (2013)
doi:10.1103/PhysRevC.88.044910
[arXiv:1303.0737 [hep-ex]].
%%CITATION = doi:10.1103/PhysRevC.88.044910;%%
%506 citations counted in INSPIRE as of 15 Feb 2020

%\cite{Chatrchyan:2012qb}
\bibitem{Chatrchyan:2012qb} 
S.~Chatrchyan {\it et al.} [CMS Collaboration],
%``Study of the Inclusive Production of Charged Pions, Kaons, and Protons in $pp$ Collisions at $\sqrt{s}=0.9$, 2.76, and 7 TeV,''
Eur.\ Phys.\ J.\ C {\bf 72}, 2164 (2012)
doi:10.1140/epjc/s10052-012-2164-1
[arXiv:1207.4724 [hep-ex]].
%%CITATION = doi:10.1140/epjc/s10052-012-2164-1;%%
%162 citations counted in INSPIRE as of 15 Feb 2020

%\cite{Wei:2015oha}
\bibitem{Wei:2015oha} 
H.~R.~Wei, F.~H.~Liu and R.~A.~Lacey,
%``Disentangling random thermal motion of particles and collective expansion of source from transverse momentum spectra in high energy collisions,''
J.\ Phys.\ G {\bf 43}, no. 12, 125102 (2016)
doi:10.1088/0954-3899/43/12/125102
[arXiv:1509.09083 [nucl-ex]].
%%CITATION = doi:10.1088/0954-3899/43/12/125102;%%
%15 citations counted in INSPIRE as of 15 Feb 2020

%\cite{Hillmann:2018nmd}
\bibitem{Hillmann:2018nmd} 
P.~Hillmann, J.~Steinheimer and M.~Bleicher,
%``Directed, elliptic and triangular flow of protons in Au+Au reactions at 1.23 A GeV: a theoretical analysis of the recent HADES data,''
J.\ Phys.\ G {\bf 45}, no. 8, 085101 (2018)
doi:10.1088/1361-6471/aac96f
[arXiv:1802.01951 [nucl-th]].
%%CITATION = doi:10.1088/1361-6471/aac96f;%%
%2 citations counted in INSPIRE as of 14 Mar 2019

%\cite{Hillmann:2019wlt}
\bibitem{Hillmann:2019wlt}
P.~Hillmann, J.~Steinheimer, T.~Reichert, V.~Gaebel, M.~Bleicher, S.~Sombun, C.~Herold and A.~Limphirat,
%``First, second, third and fourth flow harmonics of deuterons and protons in Au+Au reactions at 1.23 A GeV,''
J. Phys. G \textbf{47}, no.5, 055101 (2020)
doi:10.1088/1361-6471/ab6fcf
[arXiv:1907.04571 [nucl-th]].
%2 citations counted in INSPIRE as of 19 Apr 2020

%\cite{Hillmann:2019cfr}
\bibitem{Hillmann:2019cfr} 
P.~Hillmann, J.~Steinheimer, T.~Reichert, V.~Gaebel, M.~Bleicher, S.~Sombun, C.~Herold and A.~Limphirat,
%``Transverse momentum and rapidity dependence of collective flow harmonics of protons and deuterons in Au + Au reactions at 1.23 AGeV,''
Astron.\ Nachr.\  {\bf 340}, no. 9-10, 996 (2019).
doi:10.1002/asna.201913750
%%CITATION = doi:10.1002/asna.201913750;%%

%\cite{Bleicher:2002dm}
\bibitem{Bleicher:2002dm} 
M.~Bleicher and J.~Aichelin,
%``Strange resonance production: Probing chemical and thermal freezeout in relativistic heavy ion collisions,''
Phys.\ Lett.\ B {\bf 530}, 81 (2002)
doi:10.1016/S0370-2693(02)01334-5
[hep-ph/0201123].
%%CITATION = doi:10.1016/S0370-2693(02)01334-5;%%
%128 citations counted in INSPIRE as of 16 Apr 2020

%\cite{Reichert:2019lny}
\bibitem{Reichert:2019lny} 
T.~Reichert, P.~Hillmann, A.~Limphirat, C.~Herold and M.~Bleicher,
%``Delta mass shift as a thermometer of kinetic decoupling in Au + Au reactions at 1.23 AGeV,''
J.\ Phys.\ G {\bf 46}, no. 10, 105107 (2019)
doi:10.1088/1361-6471/ab34fa
[arXiv:1903.12032 [nucl-th]].
%%CITATION = doi:10.1088/1361-6471/ab34fa;%%
%1 citations counted in INSPIRE as of 10 Feb 2020

%\cite{Reichert:2019zab}
\bibitem{Reichert:2019zab} 
T.~Reichert, P.~Hillmann, A.~Limphirat, C.~Herold and M.~Bleicher,
%``A hadronic transport analysis on the Δ mass in Au + Au reactions at 1.23 AGeV,''
Astron.\ Nachr.\  {\bf 340}, no. 9-10, 1018 (2019).
doi:10.1002/asna.201913709
%%CITATION = doi:10.1002/asna.201913709;%%

%\cite{Motornenko:2019jha}
\bibitem{Motornenko:2019jha} 
A.~Motornenko, V.~Vovchenko, C.~Greiner and H.~Stoecker,
%``Kinetic freeze-out temperature from yields of short-lived resonances,''
arXiv:1908.11730 [hep-ph].
%%CITATION = ARXIV:1908.11730;%%
%2 citations counted in INSPIRE as of 13 Feb 2020

\bibitem{Kornakov:talk2019}
G.~Kornakov, ``Experimental results on hadron production'', \url{https://indico.gsi.de/event/8242/session/8/contribution/10/material/slides/0.pdf}

%\cite{Gossiaux:2008jv}
\bibitem{Gossiaux:2008jv} 
P.~B.~Gossiaux and J.~Aichelin,
%``Towards an understanding of the RHIC single electron data,''
Phys.\ Rev.\ C {\bf 78}, 014904 (2008)
doi:10.1103/PhysRevC.78.014904
[arXiv:0802.2525 [hep-ph]].
%%CITATION = doi:10.1103/PhysRevC.78.014904;%%
%214 citations counted in INSPIRE as of 14 Apr 2020

%\cite{Larionov:1999iw}
\bibitem{Larionov:1999iw} 
A.~B.~Larionov, W.~Cassing, M.~Effenberger and U.~Mosel,
%``(p, pi+-) correlations in central heavy ion collisions at 1 - 2 A-GeV,''
Eur.\ Phys.\ J.\ A {\bf 7}, 507 (2000)
doi:10.1007/s100500050424, 10.1007/PL00013652
[nucl-th/9910047].
%%CITATION = doi:10.1007/s100500050424, 10.1007/PL00013652;%%
%7 citations counted in INSPIRE as of 21 Apr 2020
	
%\cite{Hong:1997ka}
\bibitem{Hong:1997ka} 
B.~Hong {\it et al.} [FOPI Collaboration],
%``Abundance of Delta resonances in Ni-58 + Ni-58 collisions between 1-A/GeV and 2-A/GeV,''
Phys.\ Lett.\ B {\bf 407}, 115 (1997)
doi:10.1016/S0370-2693(97)00707-7
[nucl-ex/9706001].
%%CITATION = doi:10.1016/S0370-2693(97)00707-7;%%
%49 citations counted in INSPIRE as of 10 Feb 2020

%\cite{Brown:1991en}
\bibitem{Brown:1991en} 
G.~E.~Brown, J.~Stachel and G.~M.~Welke,
%``Pions from resonance decay in Brookhaven relativistic heavy ion collisions,''
Phys.\ Lett.\ B {\bf 253}, 19 (1991).
doi:10.1016/0370-2693(91)91356-Z
%%CITATION = doi:10.1016/0370-2693(91)91356-Z;%%
%82 citations counted in INSPIRE as of 11 Feb 2020

%\cite{Barrette:1992kh}
\bibitem{Barrette:1992kh} 
J.~Barrette {\it et al.} [E814 Collaboration],
%``Charged particle multiplicity in Si-28 + Al, Cu, and Pb reactions at E(lab) = 14.6-GeV/nucleon,''
Phys.\ Rev.\ C {\bf 46}, 312 (1992).
doi:10.1103/PhysRevC.46.312
%%CITATION = doi:10.1103/PhysRevC.46.312;%%
%26 citations counted in INSPIRE as of 10 Feb 2020

%\cite{BraunMunzinger:1994xr}
\bibitem{BraunMunzinger:1994xr} 
P.~Braun-Munzinger, J.~Stachel, J.~P.~Wessels and N.~Xu,
%``Thermal equilibration and expansion in nucleus-nucleus collisions at the AGS,''
Phys.\ Lett.\ B {\bf 344}, 43 (1995)
doi:10.1016/0370-2693(94)01534-J
[nucl-th/9410026].
%%CITATION = doi:10.1016/0370-2693(94)01534-J;%%
%561 citations counted in INSPIRE as of 14 Feb 2020

%\cite{Rode:2018hlj}
\bibitem{Rode:2018hlj} 
S.~P.~Rode, P.~P.~Bhaduri, A.~Jaiswal and A.~Roy,
%``Kinetic freeze-out conditions in nuclear collisions with $2A$ - $158A$ GeV beam energy within a non-boost-invariant blast-wave model,''
Phys.\ Rev.\ C {\bf 98}, no. 2, 024907 (2018)
doi:10.1103/PhysRevC.98.024907
[arXiv:1805.11463 [nucl-th]].
%%CITATION = doi:10.1103/PhysRevC.98.024907;%%
%4 citations counted in INSPIRE as of 14 Feb 2020

%\cite{Bass:1998ca}
\bibitem{Bass:1998ca} 
S.~A.~Bass {\it et al.},
%``Microscopic models for ultrarelativistic heavy ion collisions,''
Prog.\ Part.\ Nucl.\ Phys.\  {\bf 41}, 255 (1998)
[Prog.\ Part.\ Nucl.\ Phys.\  {\bf 41}, 225 (1998)]
doi:10.1016/S0146-6410(98)00058-1
[nucl-th/9803035].
%%CITATION = doi:10.1016/S0146-6410(98)00058-1;%%
%1398 citations counted in INSPIRE as of 14 Mar 2019

%\cite{Bleicher:1999xi}
\bibitem{Bleicher:1999xi} 
M.~Bleicher {\it et al.},
%``Relativistic hadron hadron collisions in the ultrarelativistic quantum molecular dynamics model,''
J.\ Phys.\ G {\bf 25}, 1859 (1999)
doi:10.1088/0954-3899/25/9/308
[hep-ph/9909407].
%%CITATION = doi:10.1088/0954-3899/25/9/308;%%
%991 citations counted in INSPIRE as of 14 Mar 2019

%\cite{Steinheimer:2015sha}
\bibitem{Steinheimer:2015sha} 
J.~Steinheimer and M.~Bleicher,
%``Sub-threshold $\phi$ and $\Xi^-$ production by high mass resonances with UrQMD,''
J.\ Phys.\ G {\bf 43}, no. 1, 015104 (2016)
doi:10.1088/0954-3899/43/1/015104
[arXiv:1503.07305 [nucl-th]].
%%CITATION = doi:10.1088/0954-3899/43/1/015104;%%
%26 citations counted in INSPIRE as of 14 Mar 2019

%\cite{Sombun:2018yqh}
\bibitem{Sombun:2018yqh} 
S.~Sombun, K.~Tomuang, A.~Limphirat, P.~Hillmann, C.~Herold, J.~Steinheimer, Y.~Yan and M.~Bleicher,
%``Deuteron production from phase-space coalescence in the UrQMD approach,''
Phys.\ Rev.\ C {\bf 99}, no. 1, 014901 (2019)
doi:10.1103/PhysRevC.99.014901
[arXiv:1805.11509 [nucl-th]].
%%CITATION = doi:10.1103/PhysRevC.99.014901;%%
%11 citations counted in INSPIRE as of 14 Apr 2020

%\cite{Endres:2015fna}
\bibitem{Endres:2015fna} 
S.~Endres, H.~van Hees, J.~Weil and M.~Bleicher,
%``Dilepton production and reaction dynamics in heavy-ion collisions at SIS energies from coarse-grained transport simulations,''
Phys.\ Rev.\ C {\bf 92}, no. 1, 014911 (2015)
doi:10.1103/PhysRevC.92.014911
[arXiv:1505.06131 [nucl-th]].
%%CITATION = doi:10.1103/PhysRevC.92.014911;%%
%27 citations counted in INSPIRE as of 14 Mar 2019

%\cite{Hartnack:1997ez}
\bibitem{Hartnack:1997ez} 
C.~Hartnack, R.~K.~Puri, J.~Aichelin, J.~Konopka, S.~A.~Bass, H.~Stoecker and W.~Greiner,
%``Modeling the many body dynamics of heavy ion collisions: Present status and future perspective,''
Eur.\ Phys.\ J.\ A {\bf 1}, 151 (1998)
doi:10.1007/s100500050045
[nucl-th/9811015].
%%CITATION = doi:10.1007/s100500050045;%%
%536 citations counted in INSPIRE as of 21 Apr 2020

%\cite{Ko:1999qh}
\bibitem{Ko:1999qh} 
C.~M.~Ko,
%``Description of heavy ion collisions,''
Prog.\ Part.\ Nucl.\ Phys.\  {\bf 42}, 109 (1999).
doi:10.1016/S0146-6410(99)00065-4
%%CITATION = doi:10.1016/S0146-6410(99)00065-4;%%
%1 citations counted in INSPIRE as of 21 Apr 2020

%\cite{Larionov:2001va}
\bibitem{Larionov:2001va} 
A.~B.~Larionov, W.~Cassing, S.~Leupold and U.~Mosel,
%``Quenching of resonance production in heavy ion collisions at 1 to 2 A GeV,''
Nucl.\ Phys.\ A {\bf 696}, 747 (2001)
doi:10.1016/S0375-9474(01)01216-7
[nucl-th/0103019].
%%CITATION = doi:10.1016/S0375-9474(01)01216-7;%%
%25 citations counted in INSPIRE as of 21 Apr 2020

%\cite{Steinheimer:2016vzu}
\bibitem{Steinheimer:2016vzu} 
J.~Steinheimer, M.~Lorenz, F.~Becattini, R.~Stock and M.~Bleicher,
%``Heavy baryonic resonances, multistrange hadrons, and equilibration at energies available at the GSI Schwerionensynchrotron, SIS18,''
Phys.\ Rev.\ C {\bf 93}, no. 6, 064908 (2016)
doi:10.1103/PhysRevC.93.064908
[arXiv:1603.02051 [nucl-th]].
%%CITATION = doi:10.1103/PhysRevC.93.064908;%%
%11 citations counted in INSPIRE as of 14 Mar 2019

%\cite{Barrette:1994kq}
\bibitem{Barrette:1994kq} 
J.~Barrette {\it et al.} [E814 Collaboration],
%``Measurement of pion enhancement at low transverse momentum and of the Delta resonance abundance in Si - nucleus collisions at AGS energy,''
Phys.\ Lett.\ B {\bf 351}, 93 (1995)
doi:10.1016/0370-2693(95)00329-J
[nucl-ex/9412002].
%%CITATION = doi:10.1016/0370-2693(95)00329-J;%%
%55 citations counted in INSPIRE as of 07 Apr 2020

%\cite{Stock:1985xe}
\bibitem{Stock:1985xe} 
R.~Stock,
%``Particle Production in High-Energy Nucleus Nucleus Collisions,''
Phys.\ Rept.\  {\bf 135}, 259 (1986).
doi:10.1016/0370-1573(86)90134-1
%%CITATION = doi:10.1016/0370-1573(86)90134-1;%%
%268 citations counted in INSPIRE as of 10 Feb 2020

%\cite{Adamczewski-Musch:2020vrg}
\bibitem{Adamczewski-Musch:2020vrg}
J.~Adamczewski-Musch \textit{et al.} [HADES],
%``Charged pion production in $\mathbf{Au+Au}$ collisions at $\mathbf{\sqrt{s_{NN}}}$ = 2.4$\mathbf{GeV}$,''
[arXiv:2005.08774 [nucl-ex]].
%0 citations counted in INSPIRE as of 19 Aug 2020

%\cite{Tlusty:2009dk}
\bibitem{Tlusty:2009dk} 
P.~Tlusty {\it et al.} [HADES Collaboration],
%``Charged pion production in C+C and Ar+KCl collisions measured with HADES,''
arXiv:0906.2309 [nucl-ex].
%%CITATION = ARXIV:0906.2309;%%
%19 citations counted in INSPIRE as of 15 Oct 2020

%\cite{Eskef:2001qg}
\bibitem{Eskef:2001qg} 
M.~Eskef {\it et al.} [FOPI Collaboration],
%``Identification of baryon resonances in central heavy ion collisions at energies between 1-AGeV and 2-AGeV,''
Eur.\ Phys.\ J.\ A {\bf 3}, 335 (1998)
doi:10.1007/s100500050188
[nucl-ex/9809005].
%%CITATION = doi:10.1007/s100500050188;%%
%36 citations counted in INSPIRE as of 21 Apr 2020

\end{thebibliography}
\end{document}